\documentclass[useAMS,usenatbib]{mn2e}

\usepackage{epsfig}

\newcommand{\vsini}{\mbox{$v_e\,\sin\,i$}}

\newcommand{\prot}{\mbox{$P_{\mbox{rot}}$}}

\newcommand{\kmsec}{\,\mbox{$\mbox{km}\,\mbox{s}^{-1}$}}
\newcommand{\degrees}{\mbox{$^\circ$}}
\newcommand{\rstar}{\,\mbox{$\mbox{R}_*$}}

\newcommand{\rasun}{\,\mbox{$\mbox{R}_{\odot}$}}

\title[Spot patterns  and differential rotation in the eclipsing pre-CV binary, V471 Tau]
      {Spot patterns  and differential rotation in the eclipsing pre-CV binary, V471 Tau}
\author[G. A. J. Hussain, C. Allende Prieto,  S. H. Saar, M. D. Still]
{G. A. J. Hussain$^{1,2}$\thanks{E-mail: gajh@st-and.ac.uk; callende@astro.as.utexas.edu; 
\break
saar@head-cfa.harvard.edu; martin.still@gsfc.nasa.gov}, 
C. Allende Prieto$^{3}$, S. H. Saar$^{4}$ \&
M. Still$^{5}$\\
$^{1}$ School of Physics and Astronomy, University of St Andrews, St Andrews, KY16 9SS, UK  \\
$^{2}$ Astrophysics Division, Research and Science Support Department of ESA, ESTEC, Postbus 299, Noordwijk, Netherlands \\
$^{3}$ McDonald Observatory and Department of Astronomy, University of Texas, Austin, TX 78712, USA \\
$^{4}$Harvard Smithsonian Center for Astrophysics, 60 Garden St., Cambridge, MA 02138, USA \\
$^{5}$South African Astronomical Observatory, PO Box 9, Observatory 7935, Cape Town, South Africa }
\begin{document}

\date{Accepted . Received ; in original form }

\pagerange{\pageref{firstpage}--\pageref{lastpage}} \pubyear{2002}

\maketitle

\label{firstpage}

\begin{abstract}
We present surface spot maps of the K2V primary star in the pre-cataclysmic variable binary
system, V471 Tau. 
The  spot maps show the presence of large high latitude spots located at the sub-white
dwarf longitude region. 
By tracking the relative movement of spot groups  over the course of four nights 
(eight rotation cycles), we measure the surface differential rotation rate of the system.
Our results reveal that the star is rotating rigidly with a surface shear rate, 
$d\Omega = 1.6 \pm 6$\,mrad\,d$^{-1}$.
The single active star AB Dor has a  similar spectral type, rotation period, and activity level 
as the K star in V471 Tau but displays much
stronger surface shear ($46 <d\Omega  < 58$\,mrad\,d$^{-1}$). 
Our results  suggest that tidal locking may  inhibit  differential
rotation; this reduced shear, however, does not affect the overall magnetic activity levels in active
K dwarfs. 
\end{abstract}


\section{Introduction}

V471 Tau is a well-studied eclipsing binary system
 consisting of a hot DA white dwarf and a cool K main sequence star; components of a 
 tidally locked post-common-envelope binary system (Paczynski 1976).
Photometric and spectroscopic studies of this system have revealed that 
V471 Tau is a member of the Hyades cluster at a distance of 
49\,pc and an age of approximately 500\,Myr (Nelson \& Young 1970, Vandenberg \& Bridges 1984, 
Bois, Lanning \& Mochnacki 1988, Barstow et al. 1997).
Its evolutionary status suggests that it is probably a precursor  of cataclysmic variables (CVs);
more specifically, a DQ Her-type cataclysmic variable (Paczynski 1976, Sion et al. 1998). 
Schreiber and G\"ansicke (2003) estimate that mass transfer will begin in $7.7 \times 10^8$ years on this system.
As V471 Tau is an eclipsing binary system with  K dwarf and white dwarf components, 
its photometric lightcurves can be used to obtain  precise measurements of its rotation period.
Photometry spanning over 30 years shows that the system's orbital period has varied  by 
in a quasi-sinusoidal way with a semi-amplitude of $150$\,seconds. 
The causes of these period variations remain unknown but possible explanations
include the following: (a) 
the presence of a third low mass (brown dwarf) component in the system
(Guinan \& Ribas 2001); (b)
 changes in the  system's angular momentum distribution caused by 
an exchange between magnetic and kinetic energy as
magnetic flux levels vary in a star over the course of a stellar activity cycle 
(Applegate 1988); 
or (c)  variations in the K star's rotation rate caused by mass loss from 
 a magnetically driven stellar wind. 
Guinan \& Ribas (2001) present strong evidence for the presence of a brown dwarf
companion, however after considering the third body perturbations
there are still intrinsic variations in the system's orbital period that are 
likely to be related to changes in  the magnetic activity level of the  active K2 star
 (Skillman \& Patterson 1991).

Young, rapidly rotating low mass stars (e.g. in the Pleiades cluster)
tend to show the strongest signs of magnetic activity. In the older
Hyades cluster, single  K stars  rotate relatively slowly and thus do not display strong signs of 
magnetic activity. 
Due to tidal locking with its white dwarf companion, the K star in V471 Tau is forced to rotate
at almost 50 times the solar rotation rate (\prot=0.521\,d); 
thus causing this star  to be strongly magnetically active.
Transient waves in photometric lightcurves of V471 Tau  indicate
 the presence of dark starspots  rotating around
the stellar surface (e.g. Ibanoglu 1978, Evren et al. 1986, Skillman \& Patterson 1988).
Long-term changes in the brightness level of the K star indicate a
 spot cycle with estimates of the activity cycle period, $18<P_{\rm cyc}< 20$\,yr
(Evren et al. 1986, J\"arvinen et al. 2005); although whether this variation
is truly cyclic can only be established with further observations.
There tend to be fewer large spots when the K star is brighter,
as evidenced by long-term photometry  (Fig. 2 in Skillman \& Patterson 1988).
Magnetic activity studies of V471 Tau show strong emission in 
 H$\alpha$  and Ca II H\&K  when the K star undergoes eclipse
(i.e. when the  region of the K star pointing  at the white dwarf is facing the observer), 
subsequently going into absorption  during the white dwarf eclipse 
(Rottler et al. 2002; Skillman \& Patterson 1988). 
Rottler et al. (2002) showed that the chromospheric emission diminished gradually and 
disappeared altogether between the years 1985-1992, 
implying that this phenomenon is of magnetic origin and not due to chromospheric 
reprocessing of ultraviolet (UV) radiation from the 
white dwarf as previously postulated (e.g. by Young, Skumanich \& Paylor 1998, Skillman \& Patterson 1988). 

 The white dwarf companion has an effective temperature of between 32\,000 and 35\,000\,K 
and thus contributes strongly at  X-ray ($\leq 0.3$\,keV), UV and extreme ultraviolet (EUV) wavelengths  
(Barstow et al. 1997, Guinan \& Ribas 2001 ). 
Periodic oscillations in the  amplitude of soft X-ray flux from {\sc exosat} observations
reveal that the white dwarf's photosphere is inhomogeneous and has a  9.25\,minute rotation period 
(Jensen et al. 1986, Sion et al. 1998). 
O'Brien, Bond \& Sion (2001)  trace the radial velocity variations 
of the white dwarf companion from {\em Hubble Space Telescope/Goddard High Resolution Spectrograph} (HST/GHRS) 
spectra and use these to evaluate key system
parameters, e.g. the sizes of the component stars and the inclination angle;  
they confirm that the K star is 18\% larger than other Hyades K dwarfs 
but has not yet filled  its Roche lobe.
 Sion et al. (1998) report a tentative detection of Zeeman splitting in the  HST/GHRS  Si {\sc iii} $\lambda$ 
1206 line, implying a  polar magnetic field strength of 350\,kG.

Strong absorption dips are detected in {\sc exosat} and {\sc IUE} observations of the system; 
these phenomena are attributed to  
``cool'' ($10^4-10^5$\,K) material in the inter-binary region absorbing the X-ray and 
UV radiation from the white dwarf (Jensen et al. 1986, Kim \& Walter 1998, Walter 2004).
This interbinary material is found to extend up to 2--3\rstar\  from the K star component and is likely associated with the 
3-dimensional magnetic field topology of this low mass star.
Bond et al. (2001) report the detection of coronal mass ejections (CMEs) in the V471 Tau system, which are observed as absorption transients
in the Si{\sc iii} 1206\AA\ line; this material  is ejected by the  K2 dwarf and  
 subsequently accreted onto the white dwarf (causing
the Si spot observed by Sion et al. 1998). The CME material has similar densities and masses as those
observed on the Sun but CME events are found to
 occur much more frequently, approximately 100 times more frequently than solar CME events.
Further mass loss from the K2 star occurs due to the action of a cool 
($\sim 10^4$\,K) stellar wind (Mullan et al. 1989). 
However, the white dwarf's strong magnetic field
causes a ``propeller mechanism'' which inhibits the efficiency with which material can be accreted onto the white dwarf 
(Stanghellini, Starrfield \& Cox 1990; Mullan et al. 1991).

AB Dor is an extremely well-studied single K star (\prot$=0.51$\,d, age $\approx 30$\,Myr)
(Innis et al. 1988, Collier Cameron \& Foing 1997). 
Long-term photometry of AB Dor strongly suggest the presence of a 18-20\,yr spot activity cycle, over which the
overall brightness level of the star has varied by 0.2 magnitudes (J\"arvinen et al. 2005).
However, whether this variability is truly cyclic can only be established through regular photometric monitoring
of the system over the next decade.
Spot maps of AB Dor (derived using the technique of Doppler imaging) obtained for over a decade consistently 
show the presence of a large polar cap extending from the poles to below 70\degrees\ latitude; this co-exists
with low latitude spots near 30\degrees\ latitude.
V471 Tau's K star component is similar in both spectral type and rotation rate
to AB Dor; thus by comparing surface spot maps of the two systems we can learn how 
stellar age, binarity and the resulting physical changed may
 affect the magnetic activity properties of cool stars.

Ramseyer, Hatzes \& Jablonski (1995) conducted the first  
Doppler imaging (spot mapping) study of V471 Tau's K star component, obtaining four
separate maps spanning a period of over a year.
Their spot maps show the presence of high to mid latitude spots, with a stable low latitude spot
 at the sub-white dwarf longitude.  There is  little evidence of a complete polar cap, though
possible reasons for this are discussed later in this paper.
We have mapped the surface magnetic activity patterns 
on the K star of V471 Tau in greater detail than  
has previously been possible by using new signal enhancing techniques. We exploit this higher
spatial resolution to track spot signatures over eight 
rotation periods, and thus to measure   surface differential rotation  
on the K star for the first time. We compare 
the spot patterns obtained for this K star with those  of 
the single K star AB Dor,  thus investigating how different stellar parameters affect 
magnetic activity in cool stars.

\section{Observations}

\begin{table*}
 \centering
  \caption{Journal of observations. The first three columns list the date, 
MJD times of each observation, the MJD time corrected for TT and 
heliocentric corrections, the range of phases covered, 
the name of the target (+ spectral types of standard calibration stars), 
the duration of the exposure, the number of exposures taken for each set 
of V471 Tau observations, the input S:N and the output S:N estimates. }
  \begin{tabular}{@{}llllllll@{}}
  \hline
UT Date &  MJD (TT)               & Phase        &  Target &$T_{\rm exp}$ (s)&$n_{\rm exp}$& Input S:N &	Output S:N  \\
\hline
2002 Nov 23  & 52601.1585--52601.4415 & 0.892--0.436 &  V471 Tau & 500 &    31 & 78--104 &2010--2900\\
2002 Nov 24  & 52602.1008--52602.4208 & 0.701--0.315 &  V471 Tau & 500 &    19 & 65--101 &2050--2900\\
        &                        &              &  HD 26965 (K1)& 120 & 3 & --      &  --  \\
        &                        &              &  GJ 229 (M1)  & 500 & 3 & --      &  --  \\
        &                        &              &  GJ 447 (M4)  & 900 & 2 & --      &  --  \\
2002 Nov 25  & 52603.1059--52603.2637 & 0.629--0.932 &  V471 Tau     & 500 & 12& 35--80  &1500--2550\\
        &                        &              &  GJ 860a (M3) & 900 & 2 & --      & -- \\
2002 Nov 26  & 52604.1065--52604.4290 & 0.549--0.168 &  V471 Tau     & 500 & 31& 55--91  &2040--2780\\
\hline
\end{tabular}
\end{table*}

Our data were acquired at the McDonald observatory over four consecutive nights from 2002 November
22--25 (UT dates: November 23--26)
using the cross-dispersed echelle spectrometer, cs2 (Tull et al. 1995), mounted at the 
Coud\'e focus of the 2.7-m Harlan J. Smith telescope. We obtained 
high resolution ($\lambda/\Delta \lambda \simeq $55\,000 at 5330\,\AA) 
spectra spanning over 6100\,\AA\ (3750--9930\,\AA)  of the eclipsing binary system, V471 Tau. 
By observing  this system over several nights we can track the relative movement of  surface spots 
as they are affected by surface flows, and in the process measure the  surface
differential rotation rate of the star.
Several  standard stars are also observed in order
to aid data reduction and to model the spot and photospheric contributions more accurately 
(see Table 1 for more details). The nights of November 22 and 25 were clear while
the weather on November 23 and 24 was more variable; indeed only 12 exposures of V471 Tau were
obtained on November 24. Seeing was variable throughout (between 1.5 and 3 arcsec ).

The data reduction procedure followed is conducted using standard
optimised extraction procedures in the Starlink {\sc echomop} package.
We use the signal enhancing technique of least squares deconvolution (LSD; Donati et al. 1997) 
to sum up the signal from 5000 photospheric lines contained in each exposure 
(Barnes et al. 1998). 
LSD assumes that all the photospheric lines have the same local line profile 
shapes: hence we produce a mean line profile 
by cross-correlating all the line profiles with a line mask constructed
using the line depths of each individual line. The potential 
 signal-to-noise ratio s(S:N) enhancement factor in the mean line profile can be up to $\sqrt N$ 
(where $N$ is the number of photospheric lines in the echellogram). However,  in practise the enhancement tends
to be considerably smaller
than this due to systematic effects; some of which can be accounted for (e.g.  
the line profiles cover a range of line depths, 
 noise levels vary across the echellogram), and some which cannot be accounted for as easily (e.g.
the effect of incomplete line lists used in the line mask). By measuring the noise levels in the continuum 
of our LSD profiles we can provide a check on the enhancement factor computed by the LSD programme.
We find that the LSD code used
overestimates the S:N level by between 15--20\% due to the above 
systematic effects.
The line mask only gives weight to medium strength and weak photospheric lines, 
and excluding strong chromospherically sensitive line profiles 
(e.g. H$\alpha$ at 6562\,\AA\ and the Na D doublet near 5890\,\AA).
Accurate continuum normalisation is an essential part of the LSD process; 
to ensure accurate continuum fits to each extracted echellogram we implement the 
following procedure (see Barnes et al. 1998 for more detail): \\
(a) We construct a median echellogram frame from all V471 Tau exposures for each
night and make a continuum fit to this frame by  fitting splines to each order. \\
(b) Any residual changes over the course of the night are adjusted for by dividing each
individual  V471 Tau exposure by the fit produced in (a) and fitting either low order
polynomial or spline fits to the residual frame. These frames  
are multiplied by the fit produced in (a) and the subsequent frame is 
the individual continuum fit to be used when processing  each stellar exposure. \\
(c) The LSD procedure uses the invidual continuum fit and a standard star continuum fit 
to normalise the continuum level prior to deconvolution.
 The standard star continuum
fit is produced by fitting splines to each order of  a slowly rotating inactive K1V
star (we use HD 26965 as the standard star).  This adjusts for any errors in the
individual continuum fits caused by line blending.
The edges of each order were clipped during the LSD process to minimise the
effect of bad continuum fits where the counts are relatively low as a result of a
steep blaze function.

The velocity resolution in the resulting deconvolved profiles is 5.5\,km\,s$^{-1}$. Over 
5000 photospheric lines were included in the LSD procedure; the 
peak input S:N levels varied between 35 to 104 and we obtain maximum output
S:N  levels of up to 3500 in the  deconvolved profiles  (Table 1).
The individual integrations for V471 Tau were 500 sec; which is roughly half the 
 exposure time used by Ramseyer, Hatzes \& Jablonski (1995).
The spatial resolution of  spot maps obtained using Doppler imaging depends on several factors:
the \vsini\ of the star, its rotation period, the spectral resolution
and the exposure time (which also depends somewhat on the star's brightness). 
Given our spectral resolution and the K star's \vsini\ ($v_{\rm K} \sin i$),
the optimum spatial resolution attainable corresponds to approximately
 4\degrees\ latitude at the equator.
Our 500\,s exposure times ensure that rotational blurring does not exceed 4\degrees, 
hence the spatial resolution of our spot maps is  twice that obtained by 
Ramseyer, Hatzes \& Jablonski (1995).
This resolution is necessary in order to measure surface differential rotation as accurately as possible.
Over four nights we obtained complete phase coverage and phase overlap covering approximately 
61\% of the  stellar surface. Due to the changing contribution to the 
 continuum  from the white dwarf
component during its eclipse, six exposures from the phases covering the white dwarf eclipse
and lasting a total of 6\% of a rotation phase were omitted from this  analysis. 
The  white dwarf contributes to the blue continuum, hence the blue profiles tend to appear more
shallow compared to red profiles outside of eclipse. This systematic effect will cause the
LSD profiles to be weakened slightly but does not affect the derived spot maps.

\begin{figure}
\epsfig{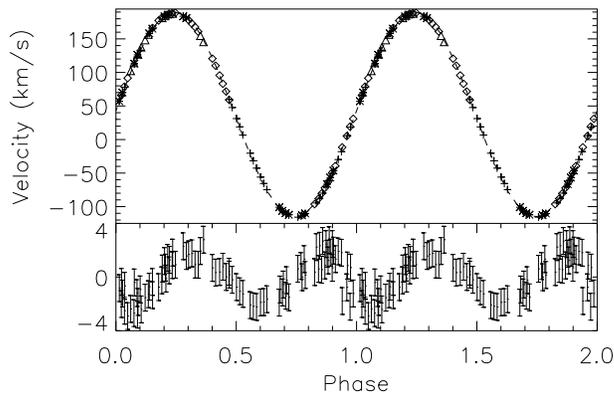}
 \caption{Top: Plot of the radial velocity curve for the K2 component of V471 Tau. 
These measurements were made by fitting a rotationally broadened profile
to the LSD profiles.
The symbols represent different nights: 22 November  (unfilled squares), 
 23  November (filled squares), 24 November (triangles) and 25 November (crosses).
Phase overlap from night to night enables us to 
track any surface shear that may occur over the course of eight rotation cycles.
The measurement errors correspond to  $pm 1.0$\,\kmsec.
The solid line represents the best fit to the radial velocity curve assuming a circular orbit.
Bottom: Residuals between the measured velocities and those computed using the best fit shown above.
}
\label{fig:radvel}
\end{figure}

\section{Optimising stellar parameters}

Initial   estimates of the basic system parameters 
(e.g. radial velocity semi-amplitude of the K star, $k_{\rm K}$, 
the radial velocity of the system, $v_{\rm rad}$, phase offset) 
can be made directly  from the LSD profiles of V471 Tau.  
These system parameters are determined to first order 
 using   an approximate rotational broadening function  to each LSD profile and 
measuring its central position in velocity-space; the rotationally broadened profile was
produced by broadening the LSD profile of a slowly rotating K1 dwarf  up to a \vsini\ 
value of 90\,\kmsec\ using the rotational broadening function (Gray 1992).
The resulting orbital velocity variations traced by the K star component are shown in Fig.\,\ref{fig:radvel}.
The error on $k_K$ as measured by least-squares fitting is approximately 0.2\,\kmsec, but 
as shown by the residuals in Fig.\,\ref{fig:radvel}, there are systematic offsets from the predicted orbital
position. Taking these systematic errors into account the overall error on $k_K$ as measured using this technique is $\pm
1$\,\kmsec.
The residual radial velocity variations are likely to be caused by inaccuracies in the velocity measurements caused by  surface spots and are in rough
agreement with the velocity perturbation by spots predicted by various models (Saar \& Donahue 1997; Hatzes 1999; 2000).
With the primary photometric perturber $\sim$60\degrees\ offset from the subobserver latitude, the Saar \& Donahue (1997) model
requires a spot inhomogeneity of $\approx$ 12\% to yield a velocity amplitude of $\pm 3$\,\kmsec. This is roughly consistent with the relative filling factor of the large spot
near longitude 315\degrees.  Thus the data demonstrate that these models are still useful even at high \vsini, and can thus be used to predict expected
velocity jitter due to spots and therefore their effect on e.g., exoplanet searches, even in very active, young, rapid rotators.  

\begin{figure}
\epsfig{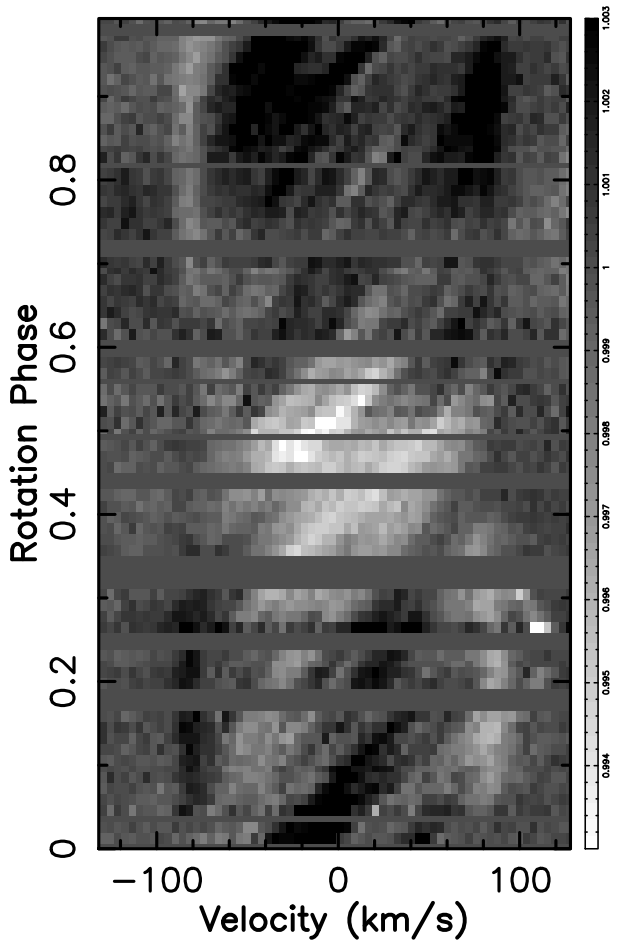}
 \caption{LSD profiles of V471 Tau, stacked
 according to the orbital phase of the system, the velocities traced in Fig.\,1 were used to correct
for the K star's radial velocity variations and the spectra were divided by the mean spectrum to 
illustrate the spot signatures more clearly. 
The white streaks mark out the movement of distortions caused by spot features moving through the line profile.
}
\label{fig:dynspec}
\end{figure}

In Fig.\,\ref{fig:dynspec} we have removed the orbital velocity variations, 
and plotted the trailed  spectrogram folded with V471 Tau's orbital period. Spots cause 
bright bumps in line profiles; by tracing the movements of these distortions across the line
profiles as the star rotates we can ascertain the latitudinal and longitudinal positions of the
surface spots and reconstruct spot maps (see review by Hussain 2004).
The bright streaks moving through the line profiles in Fig.\,2 clearly show the 
presence of surface spots. 
Phase 0.0 corresponds to the phase at which the white dwarf passes in front of the K star 
(mid-eclipse of the K star).

We use the ephemeris by Guinan and Ribas (2001), although it is modified so that 
phase 0.5 is now defined to occur when the  white dwarf companion undergoes mid-eclipse.
\[T_{0} = {\rm HJED}\, 2440610.31752 + 0.521183398E. \] 
Through least-squares fitting to the orbital velocity variations, we find the following 
 best  fit parameters: $v_{\rm rad}=34.8 \pm 1$\,\kmsec,  $k_{\rm K} =151.1 \pm 1$\,\kmsec\ 
(for the radial velocity of the system and the radial velocity semi-amplitude
of the K star   respectively). 
It should be noted that the systematic errors caused by the presence of distortions
due to starspots dominate the error estimates of $v_{\rm rad}$ and $k_{\rm K}$, 
as this method measures these quantities with a great degree of precision.
We also vary 
the zero-phase of the system to account for any variations in the ephemeris 
that may be caused by the presence of a third-body in the system. 
Table 2 shows how close these values are  to previously published values.

\begin{figure}
\epsfig{file=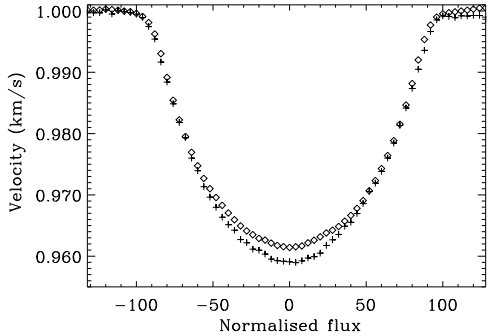,width=7cm}
 \caption{Estimating the optimum $v_{\rm K} \sin i$ value by averaging LSD line profiles.
The diamonds represent the  line profile obtained by taking the mean of all the LSD profiles
 at each  velocity bin. The crosses show the line profile obtained by taking the median of 
the lowest 10\% of all LSD profiles at each velocity bin. The mean profile is clearly filled in
due to the effect of transiting starspot distortions.}
\label{fig:avprof}
\end{figure}

An independent analysis of the optimum $v_{\rm K} \sin i$ value is made by centering the line profiles
and obtaining an average profile of the ``unspotted'' star. Doppler imaging papers 
(e.g. Ramseyer, Hatzes \& Jablonski 1995) often measure the optimum $v_{\rm K} \sin i$  value 
to be used in the Doppler imaging process by taking the mean of all the line profiles in their dataset
and fitting this mean line profile. As shown in Fig.\,\ref{fig:avprof}
 the average profile becomes ``filled-in''
by transiting spot signatures, potentially leading to systematic errors in the determination
of the final \vsini\ value. We merge all our centred LSD profiles 
(subtracting the velocity offsets of the system) and compute the median from  the 10\% 
of data points with the lowest flux 
 at each velocity bin. This effectively lets us track the least spotted
regions of the surface as they rotate through the profile and define the profile
for the purposes of deriving \vsini.
The resulting profile (Fig.\,\ref{fig:avprof}) 
enables us to obtain
a line depth closer to the ``unspotted'' level of the star and thus we can better
evaluate the  $v_{\rm K} \sin i$
value. The best fit to the line profile shown here 
is obtained with  a $v_{\rm K} \sin i=91 \pm 2 $\,\kmsec.
As Fig.\,\ref{fig:avprof} shows, this 
median profile essentially removes the artificial enhancement of 
 flux in the line core caused by transiting mid to high latitude spots. 
It is likely that this line profile is  broader than the actual  line profile due to 
uncertainties in the velocities used to centre the line profiles and remove
the orbital  variations of the K star. 

\begin{figure*}
\epsfig{file=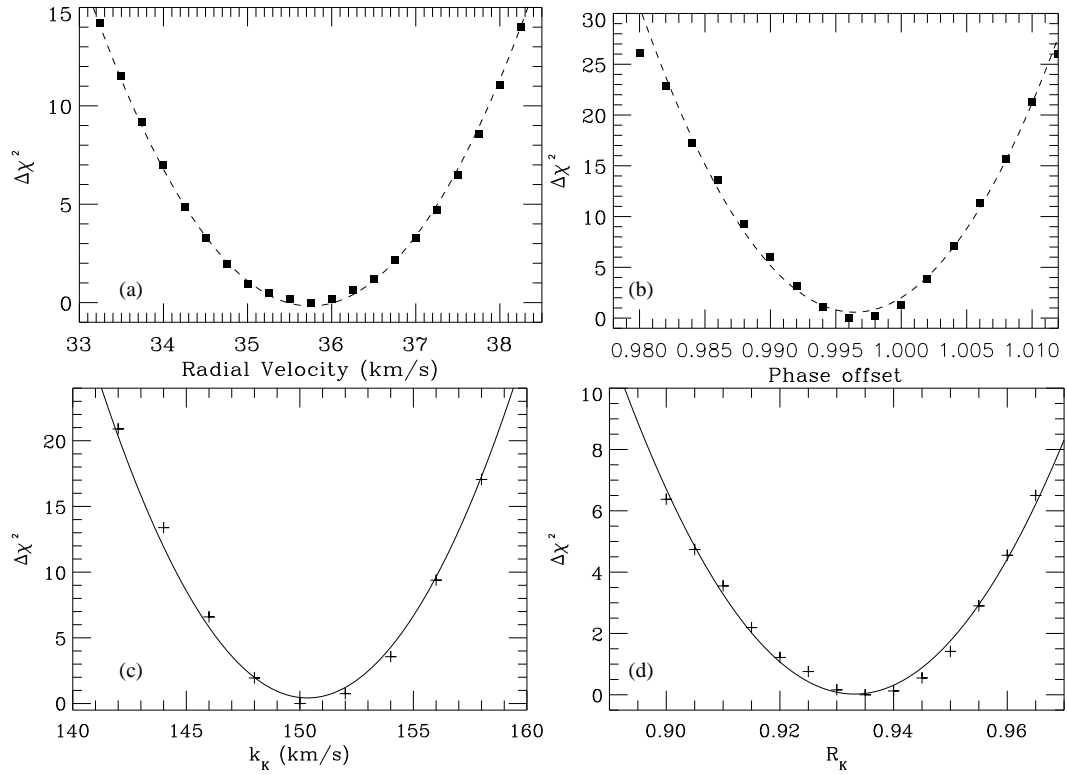, width=14cm}
 \caption{Measuring system parameters by minimising the $\chi^2$ statistic. Plots (a)-(d) show 
how the radial velocity of the system,  phase offset in the ephemeris, 
 radial velocity semi-amplitude
of the K star and the radius of the K star respectively are determined.
Parabolas are fitted in each case, the optimum value for each parameter is determined
by measuring the minimum point of these fitted parabolas. 
}
\label{fig:mins}
\end{figure*}

\subsection{Fine-tuning parameters}
The presence of large starspot signatures can affect precise measurements of the orbital velocity 
variations using the above fitting methods, so   more precise values 
of the system parameters   can be obtained within the Doppler imaging process.  Using the
Doppler imaging  code, {\sc dots} (Collier Cameron 1997), we can 
measure system parameters by optimising the level of the fit to the observed spectra 
(as measured by  $\chi^2$);  Hussain et al. (1997) and Barnes et al. (1998) discuss how  minimising
spot coverage and $\chi^2$  can be used to determine stellar parameters 
using the Doppler imaging process. 
 Some of these parameters (e.g. radial velocity and phase offset) are independent of each other and
so these determinations can be made separately
((Fig.\,\ref{fig:mins}\,a\,\&b).; 
others such as the K star radius ($R_{\rm K}$) and equivalent width (EW) are interdependent and
strongly affect the $v_{\rm K} \sin i$ value, 
thus we find the pairs of values for which  the $\chi^2$ level is minimised. 
The radial velocity semi-amplitude, 
 $k_{\rm K}$, also changes the line profile shape slightly 
(through changing the amount of velocity shear over the course of an exposure);
and then once the optimum 
$R_{\rm K}$/EW pair of values are determined, we carried out the $\chi^2$ minimisation process
again finding that the  optimum $k_{\rm K}$ value remains unchanged.

The final system parameters determined using these methods are
compared with previously published values in Table 2, and the 
agreement is generally very good. 
We measure a slightly lower value for 
$R_{\rm K}$, and therefore the  $v_{\rm K} \sin i $ value, compared to previous measurements
(these are also dependent on the EW value, see Fig.\,\ref{fig:mins}d). The radius, $R_{\rm K}=0.94$\,R$_{\odot}$
is consistent with a K dwarf that fills approximately 70\% $\pm 2$\% of its Roche lobe 
(computed from O'Brien, Bond \& Sion 2001).
The largest contribution to the error on the $v_{\rm K} \sin i$ value
is probably caused by uncertainty in the continuum normalisation process.

\begin{figure}
\epsfig{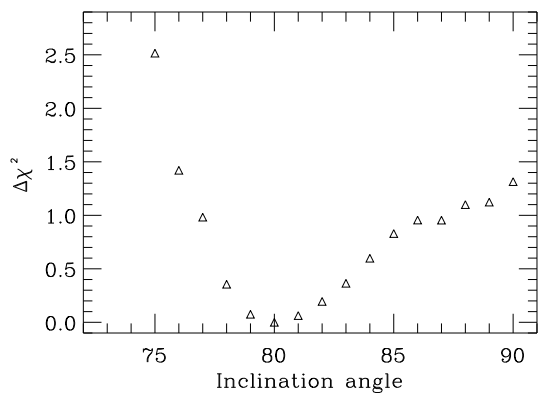}
 \caption{Measuring system parameters by minimising the $\chi^2$ statistic. 
The inclination angle of the system are measured within the Doppler imaging code.}
\label{fig:mininc}
\end{figure}

We estimate the inclination angle of the system using a similar version of the same $\chi^2$ minimisation technique
(Fig.\,\ref{fig:mininc}). While this technique is not precise, particularly at high inclination angles ($i \ge
65$), simulations conducted by Barnes (2000) indicate that the correct inclination can still be recovered at
inclination angles near 75\degrees.
As shown in Fig.\,\ref{fig:mininc},  the curve undergoes a clear minimum at
80\degrees; this is consistent with the 
value of  $i=77$\degrees$^{+7{\mbox{$^\circ$}}}_{-4{\mbox{$^\circ$}}}$ 
determined by O'Brien, Bond \& Sion (2001) and is the value we adopt in the subsequent images.

\begin{table}
 \centering
 \begin{minipage}{70mm}
  \caption{Stellar parameters: References for published values 
  (1. Bois, Lanning \& Mochnacki 1988; 2. Ramseyer et al. 1995;  3. O'Brien, Bond \& Sion 2001).
The ephemeris : $2440610.31752 + 0.521183398E$  was used for the epoch and period of system.
}
  \begin{tabular}{@{}lll@{}}
  \hline 
Parameter & Published value & Value used \\
\hline
$v_{\rm rad}$ (km\,s$^{-1}$)      &	37.4 (1)	&   35.7 $\pm 0.8$\\
$v_{\rm K} \sin i$ (km\,s$^{-1}$)& 	91$\pm 4$ (2) & 	89.5$\pm 2$		\\
$k_{\rm K}$ (km\,s$^{-1}$)       & 	148.5$\pm 0.5$ (1)	&   150.4$\pm 2.0$  \\
phase offset 	& 	--	  & 	-0.0035$\pm 0.002$\,$P_{\rm rot}$\\
$R_{\rm K}$  ($\mbox{R}_{\odot}$)& 	0.96$\pm 0.04$ (3)	&   $0.94 \pm 0.02$ \\
$i$	&   $77^{+7}_{-4}$\degrees\	(3)	& $80^{+7}_{-3}$\degrees\ \\
\hline
\end{tabular}
\end{minipage}
\end{table}

\section{Doppler imaging: spot maps from 22 \& 25 November 2002}

\begin{figure*}
\epsfig{file=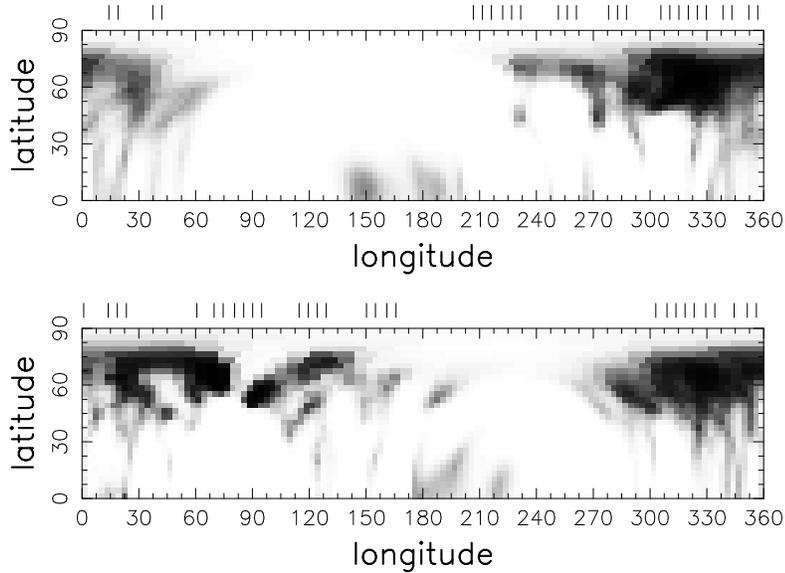, width=11cm}
\caption{ The Doppler maps from November 22 ({\em top}) and 25 ({\em bottom}) 
respectively. These maps are Cartesian projections of the stellar surface, with black representing
complete spot coverage and white representing the unspotted photosphere.
As the stellar inclination is close to 90\degrees\ spot features are mirrored in the 
``lower'' hemisphere, so only one hemisphere is shown in these plots.
The tick marks above each plot indicate the phases of observation; 
65\% of the stellar surface can be observed over the course of one night.
}
\label{fig:maps}
\end{figure*}

The Doppler imaging code used has been discussed in detail elsewhere
(e.g. Collier Cameron 1997, Hussain et al. 2000). 
On determining the best system parameters we produce spot maps, initially assuming
rigid rotation (i.e. no differential rotation). The resulting maps for 2002 November 22 and 25 
are shown in Fig.\,\ref{fig:maps} and the fits correspond to a reduced $\chi^2=1.5$.
The co-ordinate system defines  0\degrees\ longitude to occur at phase 0.0
(i.e. the sub-white-dwarf longitude or inner Lagrangian point), 
180\degrees\ longitude at phase 0.5, and longitudes 270\degrees\ and 90\degrees\ occur at phases 0.25 and 0.75
respectively.  Observation phases are denoted
by tick marks above each spot map.
These maps  show the spot coverage levels quantified by a
 filling factor that is defined as the fraction of each image pixel that is covered with spots; the filling factor
 ranges from 0 -- in the unspotted photosphere -- to 1.0 -- when the pixel has 100\% spot coverage.
These maps show  spot filling factors ranging between 8--9\% for these maps;  these values 
are typical for Doppler maps active cool stars. The spot coverage measurements
derived from the spot maps are  lower limits  as
Doppler imaging is not sensitive to spots below the resolution limit of the maps
(4\degrees\ latitude at the equator).

Our maps compare well with those reconstructed  by Ramseyer, Hatzes \& Jablonski (1995).
When comparing these Doppler maps to those from other systems 
it should be noted that the Doppler imaging process is likely to underestimate the contribution
from polar spots in high inclination stars like V471 Tau (as latitudes greater than
80\degrees\ have very little effect on the shape of the line profile). Another systematic 
effect regarding gaps in phase coverage 
when Doppler imaging high inclination stars is that very little information is
recovered at high latitudes ($>60$\degrees) 
more than $15$\degrees\ longitude away from the last phase of observation preceding the gap.
In stars with lower inclination angles, $30$\degrees$<i<60$\degrees, one can  observe
``over'' the pole and so it is possible to obtain more longitudinal coverage even with  phase gaps.
Taking these limitations into account it is apparent that the spot distribution on 
V471 Tau is very similar to those on other rapidly rotating K stars, with a distribution of
both high and low latitude spots.

As mentioned earlier the K stars in AB Dor and V471 Tau have similar rotation rates and convection
zone depths; both of these stellar parameters are thought to be key quantities in determining 
stellar magnetic activity levels with simple mean-field dynamo models
predicting that the dynamo number, $N_D$,  should scale as follows: $N_D \sim R_{\circ}^{-2}$;
where $R_{\circ}$ is the Rossby number (e.g. Parker 1979).
Observations of chromospheric and coronal activity indicators such as Ca II H\&K emission and
X-ray luminosity show that
stellar magnetic activity levels rise with decreasing Rossby number, $R_{\circ}=P_{\rm rot}/\tau_c$
(here $P_{\rm rot}$ is the stellar rotation period and $\tau_c$ is the convective turnover timescale)
(e.g. Noyes, Weiss \& Vaughan 1984, Feigelson et al. 2003), until they plateau at specific Rossby numbers. 
It is unclear what causes magnetic activity indicators to plateau, though possible explanations include
the saturation of the underlying dynamo or changes  in the way magnetic energy is 
deposited throughout the stellar atmosphere in the most active stars.
This phenomenon is called saturation: both V471 Tau and AB Dor have X-ray luminosities that lie 
within this ``saturated'' regime.

AB Dor typically shows the presence of a large stable polar cap,
which extends down to below 70\degrees\ latitude, co-existing with low latitude (equatorial) spots.
The spot maps of V471 Tau's K star shown here clearly indicate the presence of spots in the 
mid to high latitude regions, but no large polar cap covering the entire polar region. This is likely
due to the lack of complete phase coverage over the course of individual nights.
As V471 Tau has a high inclination angle ($i= 80$\degrees), very few spots are 
recovered below 0\degrees\ latitude; those that are reconstructed are  artefacts caused by 
mirroring between both hemispheres.

Given V471 Tau's orbital period of 12.5\,hours, almost two-thirds of the stellar surface is covered
in one night. Observations taken over all four nights enable us to cover  missing phases.
The phase overlap obtained over this timescale allows us to track the relative movements
of spot groups from night to night. 
On comparing the areas of phase overlap in the spot maps from November 22 and 25 (Fig.\,\ref{fig:maps}),
we find that there is no evidence of flux emergence occurring over four nights
as the spot features appear unchanged. 
This is consistent with previous observations of spot properties in active single 
and binary systems (e.g. Donati et al. 1999; review by Hussain 2002);
and means that these maps can be used to measure surface differential rotation.

\subsection{Surface differential rotation}

A direct method to measure surface differential rotation on rapidly
rotating stars involves cross-correlating constant latitude slices in spot maps 
acquired several rotation cycles apart (assuming rigid rotation in each case) 
-- the relative movements of 
surface spots within the intervening time are used to track surface flows; 
(Donati \& Collier Cameron 1997;
Donati et al. 1999). There are some  disadvantages to this method including the following:
(a) that by 
  by assuming solid body rotation over each rotation cycle it is not possible to  
  account for any surface shear  that occurs over the course of each rotation cycle; and (b)
 this method can only be applied to datasets where two full images covering the same 
phases are cross-correlated (in practise maps from successive rotation cycles have different phase
coverage); 
and finally, there is no straightforward method to measure 
the uncertainty on the derived  differential rotation parameters.

A more rigorous way to measure surface differential rotation
is to incorporate  differential rotation within the Doppler imaging code and to evaluate which values of
differential rotation give the best level of agreement (as measured using  $\chi^2$; see Petit, Donati \& Collier Cameron 2002). 
As discussed by Donati, Collier Cameron \& Petit (2003), this method has the significant advantage that sparse datasets 
spanning many rotation cycles can be incorporated, and arguably more importantly, this method can be used 
to evaluate the significance level of any differential rotation measurements. 
We employ this latter method as we can obtain over 60\%  overlap in the rotation phases observed 
over the course of all four nights. Measuring differential rotation by 
cross-correlating constant latitude slices from our most
complete maps from separate nights (22 and 25 November) would not yield a reliable result as these two maps
 only have  25\% phase overlap.

\begin{figure}
\epsfig{file=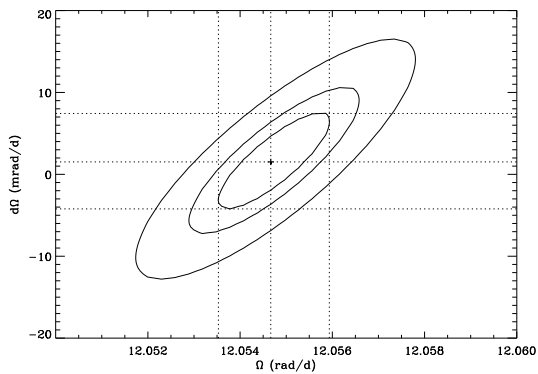, width=7cm}
\caption{The differential rotation measurement  for V471 Tau, 
the shear ($d\Omega$) and rotation rate ($\Omega_{eq}$)
values are in  units of mrad and rad respectively.
The cross marks the following best fit parameters for our dataset: $\Omega_{eq} = 12.0547\pm 0.001$\,rad\,d$^{-1}$ and
$d\Omega=1.6\pm 6$\,mrad\,d$^{-1}$
indicating solid body rotation. The contours indicate the 1, 2.67 and 5-$\sigma$ confidence 
limits on each differential rotation parameter taken individually.  V471 Tau's surface differential rotation rate is 
considerably lower than that for  AB Dor 
($46 <d\Omega < 58$\,mrad\,d$^{-1}$).
}
\label{fig:diffrot}
\end{figure}

\begin{figure*}
\epsfig{file=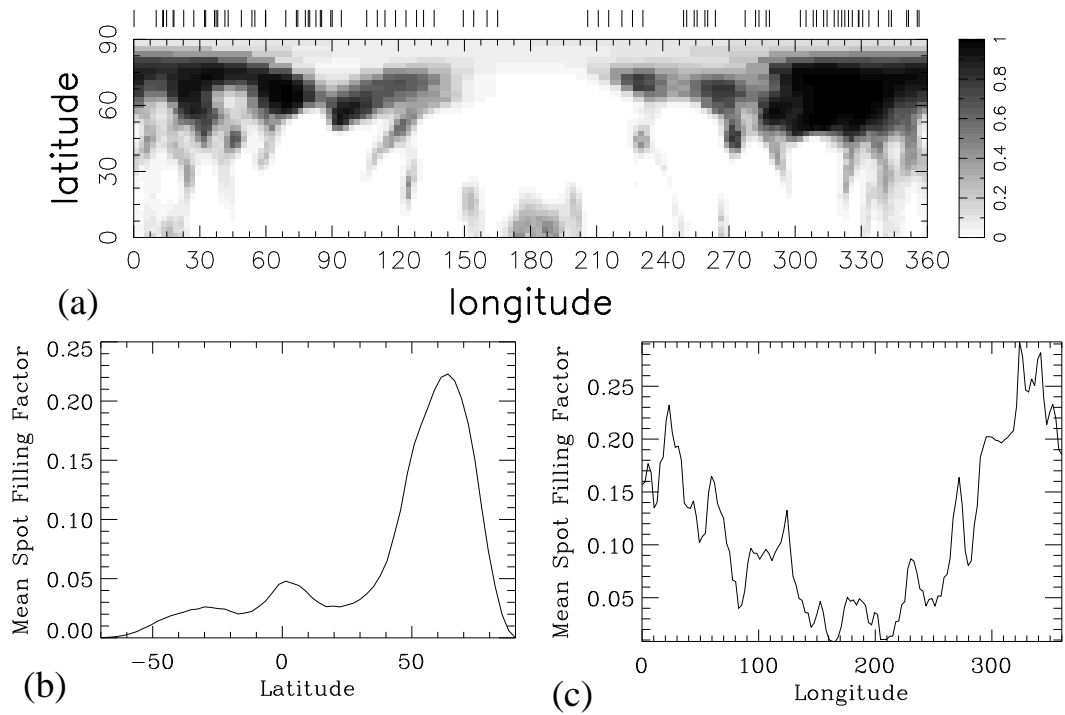, width=14cm}
\caption{(a) The combined Doppler  map from all four nights  obtained using the optimum differential rotation
parameters (as determined in Fig.\,\ref{fig:diffrot}). 
(b \& c) The mean spot filling factor as a function of latitude and longitude. The filling factor represents the percentage
of the pixel area covered by spots, with zero indicating the immaculate unspotted photosphere.
The largest spot filling factors are at the mid to high latitude regions  (55\degrees\--75\degrees\ latitude).
}
\label{fig:combimage}
\end{figure*}

Following the same procedure  described by Petit, Donati \& Collier Cameron (2002),
 we incorporate differential rotation assuming the following rotation law:
\[\Omega_{eq}(l) = \Omega_{\rm eq} - d\Omega\sin^2 l\]

Here $\Omega $ is the rotation rate at each latitude $l$, $\Omega_{\rm eq}$ is the rotation rate at the equator,  and d$\Omega$ is the difference between the rotation rates at the 
pole and the equator. Positive values of $d\Omega$ indicate solar-type rotation laws, in which the 
equator rotates more rapidly than the pole. We measured the differential rotation by
finding the pairs of $\Omega$--$d\Omega$ values for which $\chi^2$ is minimised.
Fig.\,\ref{fig:diffrot} shows a map of the resulting level of agreement  ($\chi^2$)
as a function of both differential rotation parameters. 
The cross marks the best-fit parameters to our dataset and the 
contours correspond to $\Delta \chi^2= 1$, 2.30 and 4.6:
 corresponding to 68\%, 90\%, and 99\% confidence levels on each differential rotation parameter taken
individually.
We find a shear value, $d\Omega = 1.6\pm 6$\,mrad\,d$^{-1}$; 
this is a much smaller rate of surface shear than that found on AB Dor and indeed any other
active single G and K dwarf stars 
(Barnes et al. 2005). 
Donati, Collier Cameron \& Petit (2003) use the same approach to measure differential rotation on 
 AB Dor, and find much higher differential rotation values with 
estimates of $46 <d\Omega <58$\,mrad\,d$^{-1}$.
There is  evidence that AB Dor's  surface differential rotation rate is  changing  from year to year; we discuss 
the possible reasons for this  in Section\,\ref{sec:discussion}.

V471 Tau's differential rotation rate is significantly slower than that  
recovered for the Sun ($d\Omega_{\odot} \simeq 49$\,mrad\,d$^{-1}$,
and for the analogous rapidly rotating main sequence K star, AB Dor.
The key difference between AB Dor and V471 Tau is the fact that the K star 
in V471 Tau is tidally locked with a white dwarf (and that it is a post-common-envelope 
binary). The lack of surface shear on V471 Tau is similar to 
that observed on the magnetically active K1 subgiant component, in the RS CVn binary, HR1099  
(\prot=2.84\,d, $-3.8<d\Omega<21$\,mrad\,d$^{-1}$; Petit et al. 2004). 
Our result would suggest that tidal locking with a binary component has a significant effect
on surface differential rotation in rapidly rotating K dwarfs.

\section{Discussion}
\label{sec:discussion}

We have produced a spot map of the K star component in the eclipsing
binary system, V471 Tau (Fig.\,\ref{fig:combimage}). The spot patterns obtained
are typical for an active G and K dwarf:
we find evidence of spots covering 
both low and high latitude regions, with the high latitude spots likely to reach up to the pole. 
Published  Doppler maps show starspot distributions that are predominantly at high latitudes
rapidly rotating  G--K-type stars, whether they are 
main sequence, pre-main sequence, sub-giant in
RS CVn binary systems  or giant FK~Com-type stars 
(see review by Hussain 2004); frequently these high latitude spots 
 form polar caps that completely cover the polar regions of the stars. 
Flux emergence models show that magnetic field can be transported
to high latitudes in rapidly rotating main sequence stars by the
deflection of  flux tubes polewards as they rise through their  convection zones  
(Sch\"ussler \& Solanki 1992, Granzer et al. 2000). 
This poleward deflection only occurs in rapidly rotating stars where Coriolis forces dominate.
In contrast, flux tubes in more slowly rotating stars like the Sun are subject to a much weaker  Coriolis effect
and buoyancy leads the flux tubes to rise in a radial 
direction causing very little flux to emerge at high latitudes. 
The formation of  polar spots and bimodal
spot distributions (high and low latitude spots) 
 are likely due to flux transport
mechanisms (e.g. meridional circulation, magnetic stresses) modifying the
distribution of flux after it first emerges (Deluca et al. 1997, Granzer et al. 2000). 
Schrijver \& Title (2001) show that polar spots can form in
stars that have active region emergence rates that are thirty times the solar rate.
In these models, magnetic fields form rings of opposite
polarity at the poles due to the combined action of
meridional flows and supergranular motion. 
Our results suggest that tidal locking does not interfere with the processes causing the formation of
high latitude/polar spots and a bimodal spot distribution (i.e. both high and low latitude spots) in 
rapidly rotating K dwarfs  like AB Dor and V471 Tau (Fig.\,\ref{fig:combimage}b). 

The peak spot coverage level in the Doppler map is offset from 0\degrees\ longitude (Fig.\,\ref{fig:combimage}(c)), although
the peak spot coverage level is somewhat dependent on the number of observations made at each phase.
A notable feature about the spot map shown in Fig.\,\ref{fig:combimage} 
is that surface spots are located at the sub-dwarf longitude (also known as 0\degrees\ longitude or inner Lagrangian point).
This spot pattern is consistent with predictions from both models of tidal forces acting on flux tubes emerging from the 
convection zone (Holzwarth \& Sch\"ussler 2003) and on the tidal enhancement of dynamo action itself (Moss, 
Piskunov \& Sokoloff 2002).
In CV binaries,  spots in the low mass accreting star
are used to explain changes in mass transfer rates, even causing  the transition between high and low accretion
states (e.g. Livio \& Pringle 1994, King \& Cannizzo 1998). Low latitude spots
near the inner Lagrangian point can inhibit accretion onto the white dwarf companion.
Our results would suggest that spots  are found at this very point 
in rapidly rotating K dwarfs. Theoretical  predictions from 
models of flux emergence  in close binary systems suggest that tidal forces exerted by  a binary companion
should affect the rise of flux tubes through the convection zones, potentially leading to spots
forming at preferred longitudes  with respect to the direction of the tidally locked
companion  (Holzwarth \& Sch\"ussler 2003). While we could not find find evidence for these preferred
longitudes (given the short timescale of our observations), spots at the inner Lagrangian point 
may cause the low accretion states  observed in CVs. These low accretion states  are found to last up to a period of weeks, consistent with 
estimates of spot lifetimes  in low mass stars from photometric and Doppler imaging techniques
(Hussain 2002).

Using data spanning  eight rotation cycles we find that the K star is essentially rotating as a solid body;
with significantly less surface shear than that found in the analogous single  star, AB Dor, 
despite the similarity in the stars' rotation rates. 
In some  flux-transport dynamo models, 
meridional flows are the key parameters governing the
timescale of an activity cycle (Babcock 1961, Leighton 1969). 
The meridional flow transit time -- the time taken to 
complete one cycle of the convection zone -- is crucial for setting the timescale
of a magnetic oscillation (or activity cycle period, $P_{\rm cyc}$) (e.g. Dikpati \& Charbonneau 1999). 
If the activity cycle lengths for both AB Dor
and V471 Tau are similar  ($18<P_{\rm cyc}<20$\,yr), this would suggest that
tidal locking from a binary companion does not affect the meridional flow transit time.
Alternatively, any inhibiting effect of surface flows exerted by a binary
may be compensated by enhanced  subsurface flows,  resulting in an unchanged
$P_{\rm cyc}$.
However, this can only be established from continued photoAmetric monitoring of both systems.
Guinan and Ribas (2001) indicate that there are low amplitude oscillations 
in V471 Tau's $O-C$ diagram that might be caused by activity cycle variations, but only continued monitoring could
establish whether these variations are periodic and enable us to measure the length of a possible activity cycle. 
As mentioned earlier long-term photometry of AB Dor spanning over 20 years
 shows a gradual rise and fall that may be attributable to
an 18-20 year activity cycle. 

We find that V471 Tau's K star component displays very little surface shear compared to the single K star, AB
Dor. Indeed V471 Tau's surface shear is  consistent with 
solid body rotation.  This is contrary to predictions made through calculations of the effect of tides in
close binary systems (Scharlemann 1982), although observations do suggest
that binary star components filling their Roche lobes do experience diminished differential rotation than single
stars (Hall 1991); V471 Tau's K star appears to fill almost 70\% of its Roche lobe, thus potentially accounting
for its suppressed surface shear rate. 

One remaining question is how directly does differential rotation affect surface magnetic
activity patterns? It is known to be an important dynamo parameter, but does it directly
affect the distribution of surface magnetic flux? It has been difficult to address this issue
directly by comparing spot maps from Doppler imaging targets as the stars analysed tend to 
vary too much in their sizes, convection zone depths
and rotation rates. By comparing AB Dor and V471 Tau directly we can begin to address this question. 
As pointed out earlier, 
the spot map we obtain for V471 Tau is quite  typical for a rapidly rotating active G and K-type star, 
suggesting that  reduced  differential rotation  does not  affect surface magnetic activity  patterns.
Indeed it also does not appear to have a significant effect on the global activity 
levels of V471 Tau further up in its stellar atmosphere.
The emission measure distribution derived from  X-ray spectra of V471 Tau indicates
that the stellar coronae of AB Dor and V471 Tau are very similar;  
Garc\'ia-Alvarez et al. (2004) find that both of these stars
have similar coronal temperatures and element abundances. 
The X-ray observations were acquired in  January and these spot maps are from  November
of the same year: assuming that V471 Tau's surface differential rotation rate has not changed
significantly over 10 months, we can conclude that  the
stellar rotation rate and convection zone depth  are 
 more significant factors than  differential rotation (surface differential rotation, at least)
in determining the global properties of coronae in active stars
in the X-ray saturated regime. Mean field dynamo models 
predict dynamo strength (and hence the mean magnetic field 
produced) is linearly proportional to $d\Omega$ (e.g. Parker 1979). Assuming that
$d\Omega$  is similarly inhibited throughout the convection zone, this 
poses a strong challenge to standard dynamo  models. 

The radius of the K star component of V471 Tau is found to be 0.94\rasun,
a value which lies well above the Hyades ZAMS (O'Brien et al. 2001).
AB Dor also lies above the ZAMS, with an even larger radius of between 0.94 and 1.1\rasun\
(Collier Cameron and Foing 1997). Given its \vsini\ value ($89 \pm 1$\,\kmsec; Donati et al. 2003) and
inclination angle ($i=60$\degrees), we can estimate a radius of 0.91\rasun.
These stars appear to be oversized at least partly due to the extreme spot spot activity on their surfaces
causing the stars to expand to maintain the stellar luminosity.

The spot map derived for V471 Tau indicates a spot filling factor of between 8--9\% 
(accounting for the phase gaps near longitude 180\degrees); this is typical for  an
 active rapidly rotating cool star. However, this spot
filling factor does not account for small unresolved spots that are undetectable using techniques such as
Doppler imaging. 
Measurements of spot coverage levels of active stars using molecular
band fitting techniques suggest that Doppler imaging may significantly
underestimate total spot coverage levels in active cool stars 
(e.g. Neff, O'Neal \& Saar 1995). A future
extension of this work will simultaneously analyze DI profiles with
spot diagnostics such as molecular bands 
to determine the true spot filling factor. We will compare the spot filling factors derived for both
V471 Tau and AB Dor to better compare the overall surface activity levels of both stars.

Applegate (1992) predict that the differential rotation rate of V471 Tau should change from year
to year coupled with changes in the K star's magnetic activity level.
Furthermore, they predict that the star's luminosity should fall as
its differential rotation rate increases as  energy is pumped into differential rotation. 
Given the 20-year modulation in $\Delta P/P = 10^{-6}$, they compute that this 
corresponds to a  differential rotation rate varying by  $\Delta \Omega/\Omega \simeq 0.0032$ level
 for V471 Tau. 
This  would require the  differential rotation rate to vary significantly   
($\Delta d\Omega \sim 39$\,mrad\,d$^{-1}$). 
Doppler imaging studies measuring differential rotation in 
 V471 Tau over the following years will reveal  whether or not this is the case. However,  we also note that the
 effectiveness of the Applegate mechanism has recently come into question (Lanza 2005).

MacKay et al. (2005) use models allowing for different   flux emergence patterns and
varying flux  transport mechanisms to  explain 
observed starspot and magnetic field patterns on active stars like AB Dor and V471 Tau. 
Zeeman Doppler imaging observations of V471 Tau, yielding the first surface magnetic field
maps of this system will enable us to learn how the surface magnetic field is distributed
in relation to the large spotted areas mapped in this paper.

\begin{figure*}
\epsfig{file=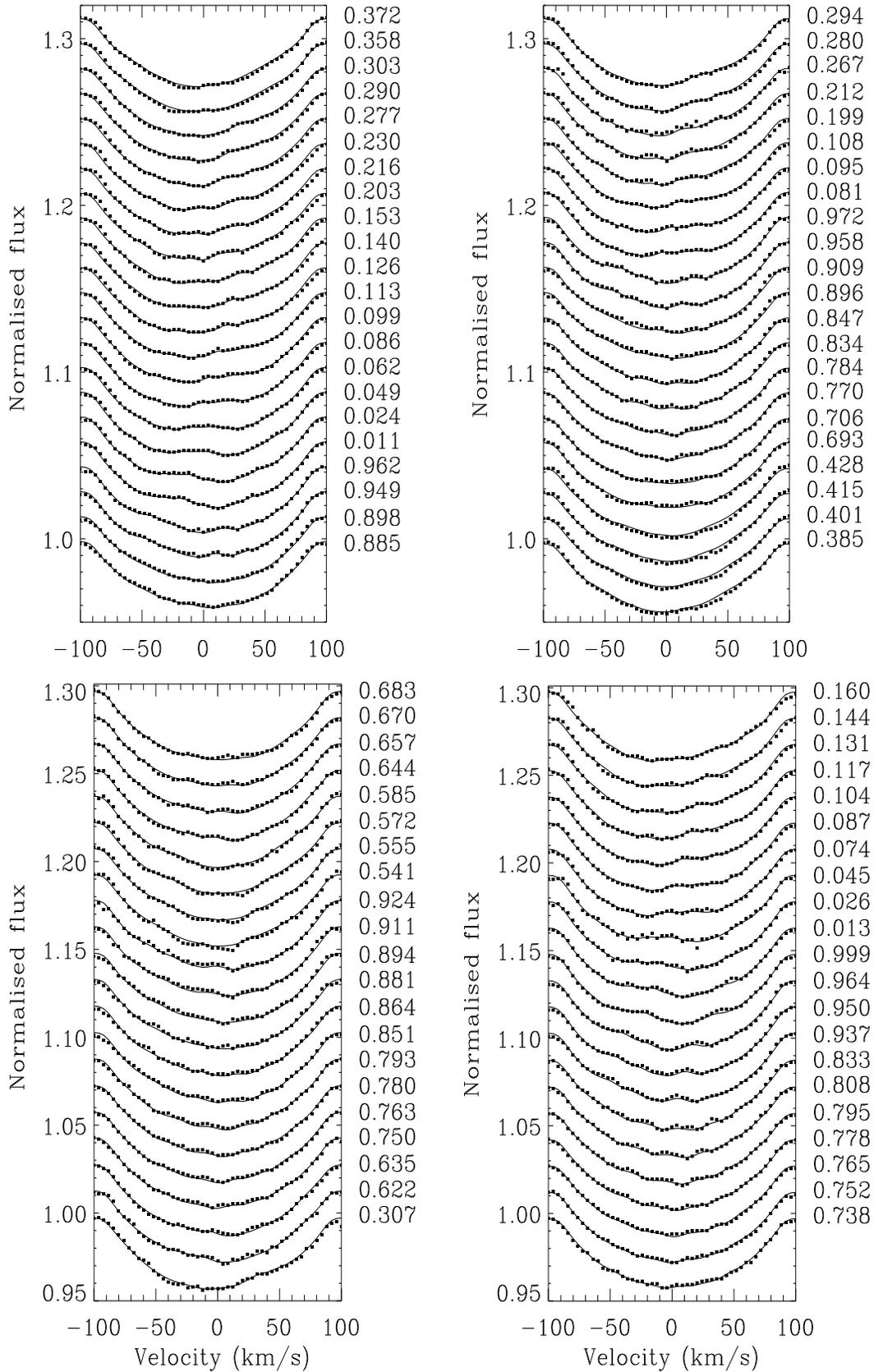, width=14cm}
\caption{Fits to the observed spectra. The  level of agreement corresponds to a reduced $\chi^2 = 1.5$.
The phase of each exposure is marked on the right hand side of each set of profiles.}
\label{fig:combfits}
\end{figure*}

\section*{Acknowledgments}
The authors would like to thank the staff at the McDonald Observatory for their help and support, especially
David Doss. We would also like to acknowledge John Barnes, Andrew Collier Cameron and the referee, Jean-Francois Donati, for
helpful comments that have improved the paper. GAJH was supported by an ESA Internal Fellowship.

\label{lastpage}


\begin{thebibliography}{99}
\bibitem[\protect\citeauthoryear{Applegate}{1992}]{applegate92} 
Applegate J.H., 1992, ApJ, 385, 621
\bibitem[]{}
Babcock H.W., 1961, ApJ, 133, 572
\bibitem[]{}
Barnes, J.R., 2000, Ph.D. Thesis, University of St Andrews, UK.
\bibitem[\protect\citeauthoryear{Barnes et al.}{1998}]{barnes98}  
Barnes J.R., Collier Cameron A., Unruh Y.C., Donati J.F., Hussain
G.A.J., 1998, MNRAS, 299, 904
\bibitem[{Barnes et al.}{2005}]{barnes05}
Barnes J. R., Cameron A. Collier, Donati J.-F., James D.J., Marsden S.C., Petit P., 2005, MNRAS, 357, 1
\bibitem[\protect\citeauthoryear{Barstow et al.}{1997}]{barstow97} 
Barstow M.A., Holberg J.B., Cruise A.M., Penny A.J., 1997, MNRAS, 290, 505
\bibitem[\protect\citeauthoryear{Bois, Lanning \& Mochnacki}{1988}]{bois88} 
Bois B., Lanning H.H., Mochnacki S.W., 1988, AJ, 96, 157
\bibitem[\protect\citeauthoryear{Bond et al.}{2001}]{bond01} 
Bond H.E., Mullan D.K., O'Brien M.S., Sion E.M., 2001, ApJ, 560, 919
\bibitem[\protect\citeauthoryear{Collier Cameron}{1999}]{cameron99} 
Collier Cameron A., 1997, MNRAS, 287, 556
\bibitem[]{cameronfoing97}
Collier Cameron A.,  Foing B., 1997, {\em Observatory}, 117, 218 
\bibitem[\protect\citeauthoryear{Collier Cameron et al.}{2001}]{cameron01diffrotabdor} 
Collier Cameron A., Donati, J.-F., 2002, MNRAS, 329, 23
\bibitem[]{}
Deluca E.E., Fan Y., Saar S.H, 1997, ApJ, 481, 369
\bibitem[]{}
Dikpati M., Charbonneau P.,  1999, ApJ, 518, 508
\bibitem[\protect\citeauthoryear{Donati \& Collier Cameron}{1997}]{donati97abdor}  
Donati J.-F, Collier Cameron A, 1997, MNRAS, 291, 1
\bibitem[\protect\citeauthoryear{Donati et al.}{1997}]{donati97lsd} 
Donati J.-F., Semel M., Carter B.D., Rees D.E., Collier Cameron A., 1997, MNRAS, 291, 658
\bibitem[]{}
Donati  J.-F., Collier Cameron  A., Hussain  G. A. J., Semel  M., 1999, MNRAS, 302, 437
\bibitem[\protect\citeauthoryear{Donati et al.}{2003}]{donati03} 
Donati J.-F., Collier Cameron A., Petit, P.,  2003, MNRAS, 345, 1187D
\bibitem[]{}
Donati, J.-F, Collier Cameron, A., Semel, M., Hussain, G.A.J., Petit, P., Carter, B.D., Marsden, S.C., 
Mengel, M., López Ariste, A. Jeffers, S.V., Rees, D.E., 2003, MNRAS, 345, 1145
\bibitem[]{}
Evren, S., Ibanoglu, C., Tunca, Z., Tumer, O., 1986, Ap\&SS, 120, 97
\bibitem[\protect\citeauthoryear{Garc\'ia-Alvarez et al.}{2004}]{garcia05} 
Garc\'ia-Alvarez D., Drake J.J., Lin L., Kashyap V.L., Ball, B., 2005, ApJ, 621, 1009
\bibitem[\protect\citeauthoryear{Granzer et al.}{2000}]{granzer00} 
Granzer T, Sch\"ussler M., Caligari P., Strassmeier K.G., 2000, A\&A, 355, 1087
\bibitem[]{gray76}
Gray D.F., 1992, The observation and analysis of stellar photospheres, Cambridge University Press, Cambridge, 
p. 374
\bibitem[\protect\citeauthoryear{Guinan \& Ribas}{2001}]{guinan01} 
Guinan E.F., Ribas I., 2001, ApJ, 546, L43
\bibitem[]{}
Hall D.S., 1991, in Tuominen I., Moss D., R{\"u}diger G., eds., Proc. IAU Coll. 130, { The Sun and Cool Stars:
Activity, Magnetism, Dynamos}, Springer-Verlag, Berlin, p.353
\bibitem[]{}
Hatzes A.P., 1999, in ASP Conf. Ser. 185: IAU Colloq. 170: Precise stellar radial velocities, p. 259
\bibitem[]{}
Hatzes A.P., 2002, {\em Astron. Nach.}, 323, 392
\bibitem[\protect\citeauthoryear{Holzwarth \& Sch\"ussler}{2003}]{holzwarth03evo2} 
Holzwarth V., Sch\"ussler M., 2003, A\&A, 405, 303
\bibitem[\protect\citeauthoryear{Hussain et al.}{1997}]{hussain97} 
Hussain G.A.J., Unruh Y. C., Collier Cameron A., 1997, MNRAS, 288, 343H
\bibitem[]{}
Hussain  G.A.J., Donati  J.-F., Collier Cameron A., Barnes  J.R.,  2000, MNRAS, 318, 961
\bibitem[\protect\citeauthoryear{Hussain}{2002}]{hussain02lifetimes} 
Hussain G.A.J., 2002, AN, 323, 349
\bibitem[]{}
Hussain G.A.J., 2004, AN, 325, 216
\bibitem[\protect\citeauthoryear{Ibanoglu}{1978}]{ibanoglu78} 
Ibanoglu C., 1978, Ap\&SS, 57, 219
\bibitem[]{innis88}
Innis J. L., Thompson K., Coates D. W., Evans T. Lloyd, 1988, MNRAS, 235, 1411
\bibitem{}
J\"arvinen S.P., Berdyugina S.V., Tuominen I., Cutispoto G., Bos. M, 2005, A\&A, 432, 657 
\bibitem[\protect\citeauthoryear{Jensen et al.}{1986}]{jensen86} 
Jensen K.A., Swank J.H., Petre R., Guinan E.F., Sion E.M., Shipman H.L., 1986, ApJ, 309, L27
\bibitem[\protect\citeauthoryear{Kim \& Walter}{1998}]{kim98} 
Kim J.S., Walter F.M, 1998, in Donahue R.A., Bookbinder J.A., eds, 
Cool Stars, Stellar Systems and the Sun, Astron. Soc. Pac. Conf. Ser., Vol. 154,  p. 1431
\bibitem[]{}
King A.R., Cannizzo J.K., 1998, ApJ, 499, 348
\bibitem[]{}
Lanza A., 2005, MNRAS, 364, 238
\bibitem[]{}
Leighton R.B., 1969, ApJ, 156, 1
\bibitem[{Livio \& Pringle}{1994}]{}
Livio M., Pringle J.E., 1994, ApJ, 427, 956
\bibitem[\protect\citeauthoryear{Mackay et al.}{2004}]{mackay04}
Mackay D.H., Jardine M., Cameron A. Collier, Donati J.-F., Hussain G.A.J., 2004, 
MNRAS, 354,  737
\bibitem[]{}
Moss D., Piskunov N., Sokoloff D., 2002, A\&A, 396, 885
\bibitem[\protect\citeauthoryear{Mullan et al.}{1989}]{mullan89} 
Mullan D.J., Sion E.M., Bruhweiler F.C., Carpenter K.G.,  1989, ApJ, 339, L33
\bibitem[\protect\citeauthoryear{Mullan et al.}{1991}]{mullan91} 
Mullan D.J., Shipman H.L., Sion E.M., MacDonald J., 1991, ApJ, 374, 707
\bibitem[{Nelson \& Young} {1970}]{nelson70}
Nelson B, Young A., 1970, PASP, 82, 699
\bibitem[]{}
Neff J.E., O'Neal D., Saar S.H., 1995, ApJ, 452, 879
\bibitem[]{}
Noyes R.W., Weiss N.O., Vaughan A.H., 1984, ApJ, 287, 769
\bibitem[\protect\citeauthoryear{O'Brien, Bond \& Sion}{2001}]{obrien01} 
O'Brien M.S., Bond H.E., Sion E.M., 2001, ApJ, 563, 971
\bibitem[\protect\citeauthoryear{Paczynski}{1976}]{paczynski76}  
Paczynski B., 1976, in Eggleton P., Mitton S., Whelan J., eds,
Proc. IAU Symp. 73, Structure and Evolution of Close Binary Systems, 
Dordrecht, p.75
\bibitem[]{}
Parker E.N., 1979,  Cosmical magnetic fields: Their origin and their activity,
 Clarendon Press, Oxford University Press, New York
\bibitem[\protect\citeauthoryear{Petit et al.}{2002}]{petit02} 
Petit P., Donati J.-F., Collier Cameron A.,  2002, MNRAS, 334, 374
\bibitem[]{}
Petit P., Donati  J.-F., Wade  G. A., Landstreet  J.D., Bagnulo  S., 
L\"uftinger  T., Sigut  T.A.A., Shorlin  S.L.S., Strasser  S., Auri\`ere  M., Oliveira J. M., 
2004, MNRAS, 348, 1175
\bibitem[\protect\citeauthoryear{Ramseyer, Hatzes \& Jablonski}{1995}]{ramseyer95} 
Ramseyer T.F., Hatzes A. P., Jablonski F., 1995,AJ,110, 1364
\bibitem[\protect\citeauthoryear{Rottler et al.}{2002}]{rottler02}  
Rottler L., Batalha C., Young A., Vogt S., 2002, A\&A, 392, 535
\bibitem[]{scharlemann82}
Saar S.H., Donahue R.A., 1997, ApJ, 485, 319
Scharlemann, E.T., ApJ, 253, 298
\bibitem[]{}
Schreiber, M.R., G\"ansicke, B.T., 2003, A\&A, 305, 321
\bibitem[{Schrijver \& Title}{2001}]{schrijver01}
Schrijver C.J., Title A.M.,  2001, ApJ, 551, 1099S
\bibitem[\protect\citeauthoryear{Sch\"ussler \& Solanki}{1992}]{schusslerflux92}  
Sch\"ussler M., Solanki S. K.,  1992, A\&A, 264, 13
\bibitem[\protect\citeauthoryear{Sion et al.}{1998}]{sion98} 
Sion E.M., Schaeffer K.G., Bond H.E., Saffer R.A., Cheng F.H., 1998, ApJ, 496, L29
\bibitem[\protect\citeauthoryear{Skillman \& Patterson}{1988}]{skillman88} 
Skillman D.R. \& Patterson J.P., 1988, AJ, 96, 976
\bibitem[\protect\citeauthoryear{Stanghellini, Starrfield \& Cox}{1990}]{stanghellini90} 
Stanghellini L., Starrfield S., Cox A.N.,  1990, A\&A, L13
\bibitem[]{tull95}
Tull R.G., MacQueen P.J., Sneden C., Lambert D.L., 1995, PASP, 107, 251
\bibitem[\protect\citeauthoryear{Vandenberg \& Bridges}{1984}]{vandenberg84}  
Vandenberg D.A., Bridges T. J.,  1984, ApJ, 278, 679V
\bibitem[]{walter04}
Walter F., 2004, AN, 325, 2411
\bibitem[]{}
Young A., Skumanich A., Paylor V., 1988, ApJ, 334, 397
\end{thebibliography}
\end{document}